# Is Semantics Physical?!

Maria K. Koleva
Institute of Catalysis, Bulgarian Academy of Science
1113 Sofia, Bulgaria
e-mail: mkoleva@bas.bg

*It is demonstrated that under the hypothesis of boundedness, the semantics appears as a property of spontaneous physical processes. It turns that both semantic structure and semantic meaning have their own physical agents each of which is represented trough generic for the state space property. The boundedness sets an exclusive two-fold representation of a semantic unit: as a specific sequence of letters and as a performance of a specific engine so that their interplay serves as grounds for building a multi-layer hierarchy of semantic structures. It is established that in this setting the semantics admits both non-extensivity, permutation sensitivity and Zipf`s law. The robustness of the hierarchical organization of semantic structures is maintained by new generic form of non-local feedback that appears as a result of the necessary for sustaining boundeness matter wave emitting.*

## Introduction

One of the most important problems in modern science is how to meet the ever-increasing demands to computing performance. So far the major concern, commonly known as Moore`s law, is that the storage capacity achieved with existing technologies will eventually reach a plateau. At present, the attention of the scientific community is focused on establishing consensus on what type of emerging technology (*quantum computers, molecular electronics, nano-electronics, optical computers and quantum-dot cellular automata*) holds most promises to keep up the current pace of progress.

Yet, the problem would be only partially set by the merits of the best technology because it is tacitly presupposed that whatever the appropriate choice would be, the computing process will follow the traditional information theory which, however, renders the computing an extensive process regardless to the technological merits of the hardware. Indeed, the extensivity of the computing commences from the assertion that every algorithm is reducible to a number of arithmetic operations, executed by means of linear processes. Thus, the more complex is the algorithm the more operations it involves and in turn, the more hardware elements (and/or time) are necessary to accomplish the task; thereby the future development of the computing performance would be reduced to the matter of compromise between the speed of operations and energy and production costs.

That is why we comprehend the major move ahead not as the best choice among the variety of contemplated future and emerging technologies but in establishing grounding principles of a next generation performance strategy that opens the door to realization of a functional circuit capable to autonomous organization of information in a hierarchy of semantic structures. The key advantage of this strategy lies in the non-extensivity of that information, a property gained from the irreducibility of a semantic content to the information stored in the sequence of symbols (units) that it comprises. The effect of the non-extensivity of information organized in semantic structures is best illustrated by the rate of it's extent – as we shall demonstrate in sec. 1.2, the latter is justified by the availability the next and the previous unit (letter, word, sentence…) to be "foreseen" on the grounds of the knowledge about the current unit only, i.e. such circuit performs as an "Oracle". To compare, according to the traditional information theory each letter can follow or precede any other one from the alphabet and thus it is impossible not only "foreseeing" the next letter, but it is impossible to judge apriori whether a given sequence is a message (i.e. it comprises information) or it is random sequence of symbols. Next we will demonstrate that this is a crucial drawback of the traditional information theory, the overcoming of which sets the grounds of "semantic" computing.

Indeed, the traditional information theory substantiates its general assertion that the information is physical and the computing is executed by natural physical processes through the suppositions that the information is to be associated with the properties of the macro states of a given physical system and the computing is to be executed by the natural motion in its state space. On the other hand, according to the traditional statistical mechanics, which is supposed to govern the behavior of the natural physical processes, all points in the state space are permanently accessible from any other point, a property well known as markovianity. This, in turn, justifies the probabilistic description of the state space motion and serves as grounds for parity between the Shannon information and the Boltzmann entropy. An immediate consequence of the above setting implies permutation invariance of the successive jumps in any given trajectory and accordingly it implies permutation invariance of the letters in a sequence. Then, indeed, in the above setting, the execution of any natural physical process does not allow distinguishing of a message from a random sequence of symbols.

On the other hand, the very idea of computing sets governing role of the logical operations over the physical properties of the hardware. The major implement for achieving this goal is hand-craft resetting of the hardware to a "neutral" state such that all information states (symbols) appear tantamount under the execution of each logical operation. It is worth noting that the reset to "neutrality" is necessary condition for execution of all numerical operations as linear processes. Yet, it is obvious that the "neutral" state is not tantamount to the information states. Thereby, a present day computer appears as an open system able to accomplish any sequence of logical operations by means of an extensive sequence of linear processes at the expense, however, that its physical properties do NOT admit traditional statistical physics.

Thus, the above considerations bring us to the dilemma whether the general assertion that information is physical and computing is executed by natural physical processes still holds and if so, at what constraints. The major goal of the present paper is to prove that under serious reconsideration of certain basic notions in information theory and in statistical physics, the assertion not only holds truly but semantics emerges naturally as well.

The first task of our program is to consider the major characteristics of semantics which we suppose to be substantiated SPONTANEOUSLY by a natural physical process. The long-standing systematic study and our daily experience select the following three ones:

(i) non-extensivity - the basic idea behind the non-extensivity of the information organized in semantic structures implies irreducibility of a semantic unit to the sequence of symbols that it comprises;

(ii) permutation sensitivity - semantic units (letters, words, sentences…) are hierarchically organized by means of syntactic and grammatical rules whose generic property is permutation sensitivity. The latter implies that semantics is to be exclusively associated with specific long-range order (correlations) among the semantic units;

(iii) probabilistic laws – given some corpus of natural language utterances, the frequency of any word is inversely proportional to its rank in the frequency table. This property is famous as Zipf`s law.

The enigmatic ubiquity of these properties along with their extreme mutual confrontation is a crucial test for our idea that the semantics is physical. Indeed, on the one hand, permutation sensitivity implies successive semantic units to appear in a specific hierarchical order; on the other hand, the Zipf`s law implies that the word order and hierarchy are ignorant. The puzzle is extremely tightened by the inability of the statistical physics to explain neither of these properties nor their coexistence and ubiquity. Moreover, markovian trajectories in the state space never close – in turn, this makes impossible to define any syntactic unit (word, sentence) because all points in the state space are tantamount in the sense that each of them belongs to one of the macro states which represent information units. In other words, there is no "space bar" to separate a given word from the next. Thus, the formation of semantic units is impossible to happen spontaneously and any if to be associated any, the latter would be a matter of "free will" of an external "mind".

The next challenge to the traditional information theory concerns the Shannon definition of information. To remind, it implies that the information is measured by the probability for executing information symbols on their random choice. Assuming validity of the law of large numbers, Shannon has assumed that this probability is proportional to the volume of the macro state which represents the information symbol considered. However, an overlooked point is that markovianity of the state space motion invalidates the law of large numbers because it sets the margins of deviation between the frequency of appearance of any information symbol and the corresponding macro state volume (probability of the event) not to converge uniformly to zero on increasing the length of the sequence. Indeed, while the macro state volumes are fixed, the trajectory can "jump" to the farthest end of the attractor by any single transition. In turn, this makes the frequency of appearance of every informational symbol wildly wandering from zero up the volume of the entire state space in a single point; thereby never converging.

Next we shall demonstrate that under the hypothesis of boundedness, introduced recently [1,2], the spontaneous motion in the state space exhibits semantics the major properties of which are non-extensivity, permutation sensitivity and Zipf`s law.

## 1. Semantics and Boundedness

### 1.1. State Space Under Boundedness

The hypothesis of boundedness [1, 2] consists of 1) a mild assumption of boundedness on the local (spatial and temporal) accumulation of matter/energy at any level of matter organization and 2) boundedness of the rate of exchange of such an accumulation with the environment. Under the assumption of the hypothesis of boundedness the state space of an open system is partitioned into basins of attraction and the trajectories form a dense transitive set of orbits. The second part of the hypothesis is implemented by restricting transitions from a point on a trajectory to being those to nearest neighbors only; i.e., transitions from any point on a trajectory must pass through an adjacent state (one can think of a latticised state space). For adjacent states [3] to be well defined, closed trajectories must belong to the same homotopy class which restricts the loci of trajectories to being simply connected. To compare, the Markovian trajectories are inherently topologically unconstrained and can therefore have as loci manifolds that are not simply connected (e.g., a torus) which in turn means that trajectories can belong to different homotopy classes.

The above difference between a bounded and a markovian motion has far-going consequences one of which is fundamental for setting semantics through natural processes. It is substantiated by the generic for the bounded motion property of self-organization into two scales – a fine-grained one which never closes and a coarse-grained one which forms a dense transitive set of orbits [2]. The condition for lack of steady fluxes sets balance between both motions [4] whose measure turns to appear as Boltzmann-Gibbs weight. The latter is a central result of our approach since, being derived free from the condition of entropy maximization, it provides:

(i) fundamental development of the idea of banning perpetuum mobile through associating of a specific engine, not necessarily physically coupled to two heat reservoirs, with each inter-basin orbit; to each engine an effective Carnot engine is assigned so that its efficiency never exceeds that of the corresponding Carnot engine [4]. Thus, the withdraw of entropy maximization as a necessary condition for achieving ban over perpetuum mobile, renders its ubiquity: now it holds not only for simple equilibrium systems (i.e., systems which admit entropy maximization) but for complex out-of-equilibrium systems, i.e., systems which apparently violate the entropy maximization as well.

(ii) as we shall demonstrate in the next sub-section at the same time it provides non-extensivity of the semantics and its permutation sensitivity.

### *1.2. Non-Extensivity of Semantics and Its Permutation Sensitivity*

The considered so far properties of the motion in the state space under boundedness are enough to outline the general implements of the semantic appearance. First, it is obvious that the boundedness over the rate of admissible transitions puts a ban over the arbitrariness of the sequences of informational symbols. Thus, it becomes transparent how one can "foresee" the next state on the grounds of the current one only – the next state is one of the admissible for the current state ones. Further, an exclusive property of boundedness is that every inter-basin orbit starts and ends at the point which represents the average of the corresponding trajectory; the existence of average for a bounded time series is guaranteed by the Lindeberg theorem [5]. Thus, the above considerations apparently suggest association of semantic units (words) with inter-basin orbits along with association of the "space bar", (i.e., the symbol which separates one semantic unit from the other) with the point that represents the "average".

It should be stressed that the above suggestion is justified by the specific structure of the state space under boundedness. Indeed, the boundedness sets as generic properties of the state space motion to be: (i) non-markovianity of the orbital motion; (ii) the "average" to be the single point in a state space where the memory is got lost. The non-markovianity of the orbital motion is a direct consequence of the restriction over the rate of transitions which makes any next transition to depend not only on the current state but on all previous ones on the orbit; by the same reason the non-markovianity selects the future states and thus makes possible to "foresee" them. On the other hand, again because of the boundedness, every orbit returns to the average, where the memory has got lost and a new orbit starts. Thus, the unique interplay between markovianity and non-markovianity makes available both formation of a different "words" and their separation one from another to happen as a natural physical process. It is worth noting that the non-Markovianity of the inter-basin motion holds even when the Chapman-Kolmogorov relation holds [5].

Further, our approach allows another generic definition of a semantic unit - through association of a word with the performance of the specific engine associated with the corresponding inter-basin orbit. This is a fundamental move ahead since it renders the non-extensivity of the semantics. Indeed, the association of any syntactic unit with an independent physical agent (corresponding engine) provides autonomous physical description of its semantic content. Indeed, physically the semantic content is measured by the useful work produced by the engine and its efficiency. Thus, the semantic content has an autonomous expression which, though interrelated, is still irreducible to the characteristics of the symbols that constitute the corresponding inter-basin orbit. Yet, as considered in sec.2.3 the two-fold representation of a semantic unit, i.e. through a specific sequence of letters and through the performance of the corresponding engine, turns central for building a many-level hierarchical structure of the semantics.

Further, the property of permutation sensitivity also naturally commences from the association of a semantic unit with the performance of a specific engine – it is provided by the functional irreversibility of any Carnot engine, i.e. the property of the latter to work as a heating machine in one direction while in the opposite one to perform as a refrigerator. Moreover, the sensitivity of the performance of an engine to the sequence of states (symbols) in the corresponding orbit renders broadening of the notion of sensitivity to permutations in the sense that it provides not only irreversibility of the word order, but change in the semantic meaning under arbitrary permutations of the letters in the matching word.

Outlining, there are two interrelated yet irreducible one to another ways of representing the non-extensivity and permutation sensitivity of a semantic unit: (i) the first one is trough the performance of the corresponding engine; (ii) the second one is trough the long-range correlations among the letters in a word set by the non-markovianity of the inter-basin motion. As we shall demonstrate in sec. 2.3 while the performance of an engine provides the meaning of a word viewed as a single unit, the long-range correlations among its letters serve as grounds for its algorithmic-like presentation, i.e. the word viewed as a correlated sequence of symbols proceeded by non-linear operations and operations that involve irrational numbers. And, the non-linear algorithmic-like representation serves as grounds for the hierarchical self-organization of the semantic structure.

It should be stressed that the autonomy of the semantics is an exclusive property of the state space motion under boundedness. To compare, since markovian trajectories never close, in the frame of the traditional statistical mechanics the semantics could appear only through apriori set decoding applied to a sequence of symbols.

The association of the semantics with the performance of an engine provides new insight on the long-standing deadlock in the discussion about the relation between the information and the Second Law [*see publications on Maxwell demon and Landauer* *principle; some of them are* 6-12]. The major move ahead is the elimination of the confrontation between "mind", viewed as a demon that exerts the non-spontaneous process of separating molecules according to their velocities, and the Second Law that governs the behavior of thermodynamically reversible, yet *spontaneous* processes. The major implement of that elimination is the association of a semantic unit with the performance of spontaneously self-organized engine. Moreover, this semantics is coherent with the Second Law because the ban over the perpetuum mobile automatically holds for the engines under boundedness.

### *1.3. Zipf`s Law and Boundedness*

Next we shall prove that the Zipf`s law arises as a consequence of the exclusive generic property of each bounded irregular sequences (BIS) which is proven to be that [1,2]: the power spectrum of any BIS is a continuous band that fits the shape $1/f^{\alpha(f)}$, where $f$ is the frequency and $\alpha(f) \to 1$ at $f \to 1/T$ ($T$ is the length of the sequence); this shape remains the same regardless to whether the terms in a BIS obey certain distribution or not; alongside, as proven in [2], this shape remains the same under any operation of coarse-graining; coarse-graining implies any non-linear operation such as local averaging, local amplification or local damping that leaves the BIS bounded. This immediately implies that for both: (i) sequences of words viewed as sequences of symbols, space bar included, and (ii) the sequences of words, viewed as semantic units, the frequency of any word is inversely proportional to its rank in the frequency table.

Thus, our task now is to show that under the constraint of boundedness, any sequence of symbols and the corresponding sequence of words (viewed as a non-uniform coarse-graining of the sequence of symbols) can be represented as a BIS. To begin with, let us remind that the association of the semantic units with the performance of the corresponding engines allows an independent representation of the symbols as the most appropriate choice of specific properties of the corresponding macro-states. Moreover, since the shape of the power spectrum of BIS remains the same regardless to whether the terms in a BIS are defined as probabilities or through other physical properties of the macro-states, each physical property of a macro-state could specify a symbol if only it is: (i) invariant under intra-basin motion but specific for inter-basin motion; (ii) remains permanently bounded. Then, since each word, viewed as an inter-basin orbit, comprises only finite numbers of letters, whatever the physical variable representing the symbols (letters) is, the deviations of that variable from the average of the corresponding time series remain bounded; alongside, the deviations remain bounded under any coarse-graining, i.e. any representation of the time series viewed as a sequence of autonomous semantic units remains also a BIS. Thereby, regardless to its semantic content, any sequence of letters (space bar included) and/or words, under the constraint of boundedness alone, appears as BIS and so obeys Zipf`s law.

Yet, the fulfillment of Zip`s law both for meaningful and meaningless sequences poses the question about the further hierarchical organization of the semantic units in the sense how a physical system and its state space should be organized so that the words to be organized in sentences in a meaningful way. In the next section we shall demonstrate that this matter is inherently interrelated with the matter about the reproducibility of the semantic order and semantic units themselves. The problem has the following two major aspects:

(i) reproducibility of a semantic unit. The problem arises from the fact that though restricted, the number of the admissible states for any given one is still more than one. Thereby, on the condition of the reproducibility of the current state, the next state is still matter of choice between the admissible ones.

(ii) generic organization of the state space able to provide meaningful organization of the words in sentences. The so far considered state space structure cannot provide further hierarchical order able to organize words in sentences. Indeed, the highly non-trivial interplay between the non-markovianity of the orbital motion and the single point where the memory is got lost (average) provides the specific semantic content of each "word" along with natural reset to the "space bar". Yet, at the same time it is not sufficient to produce the organization of the words into sentences.

In the next section we shall demonstrate how, again the boundedness, sets the general properties of a hierarchical organization capable to provide not only higher order semantical organization but its reproducibility as well.

## 2. Hierarchical Structure of Semantic Organization

### *2.1. General View Over the State Space Structure*

The generic structure of a state space able to provide higher order semantic structure must obey the following constraints:

(i) at any level of organization the semantic units (sentences) appear as dense transitive set of orbits;

(ii) the lower level semantic units (words) appear as local dense transitive set of orbits superimposed on a coarse-grained orbit (sentence) so that its local average performs as local "space bar";

(iii) no infinite length orbits are admissible.

The first two of the above constraints render the local averages to be non-markovian in the sense that the "choice" of the next word, i.e. a local choice of the next orbit, must be coherent with the exertion of the corresponding higher level orbit (sentence). Thus, the coherence between the motions on different hierarchical levels sets long-range correlations among the semantic units (e.g. words) thereby providing their organization in higher order semantic structures (e.g. sentences).

The physical implement of semantic hierarchy are spatially distributed systems subject to the following organization:

(i) they consist of nodes each of which is local out-of-equilibrium system, .i.e., a system coupled to a mass/heat reservoir and which spontaneously emits the products of its activity away from the locality where it operates. It is worth noting, that the spontaneous emitting of the products is set by the boundedness viewed as limitation over the matter/energy accumulation. In the next subsection we will demonstrate that for reaction-diffusion systems this emitting occurs in form of a specific for any given system matter wave.

(ii) the spontaneous emitting of the products provides specific driving of distant nodes. Since the matter wave associated with any spontaneous emitting is a generic property of all nodes, the successive driving of other nodes makes the set of nodes densely connected. Yet, this connectivity provides a stable steady functioning of the entire network if and only if the nodes are connected so that to perform in steady cycles. The stability of the functioning into steady cycles is provided by the exclusive property of the spontaneously emitted matter waves to damp down local fluctuations of the control parameters. Indeed, this property is due to the fine-tuning of the spontaneously emitted matter wave to the current control parameters of the corresponding node. Then, any deviation in the control parameters immediately modifies the characteristic of the matter wave; it in turn causes a cascade of changes in the entire cycle of connected nodes which turns back as modification in the driving force of the initial node. Thus, for steady cycles, the spontaneous emitting of a matter wave serves as stabilizing factor. Moreover, the fine-tuning of the matter wave to the control parameters and the subsequently caused cascade of modifications in the corresponding cycle can be considered as a generic form of non-local autocatalysis whose particular role is the stabilization of the corresponding reaction rate.

Thereby, the organization of a network into steady cycles with coherent hierarchy of their performing provides not only semantic hierarchy but the robustness of the latter under reproducibility.

## *2.2. Spontaneous Emitting of Matter Wave*

Next we shall figure out the characteristics of a spontaneously emitted matter wave. For this purpose, let us consider a very simple example of a reaction-diffusion system coupled to mass and heat reservoirs whose behavior is described by the following system of non-linear differential equations:

$$\frac{\partial \vec{X}}{\partial t} = \vec{\alpha} \bullet \hat{A}(\vec{X}) - \vec{\beta} \bullet \hat{R}(\vec{X}) - \nabla \bullet (\hat{D}(\vec{X}) \bullet \nabla \vec{X}) \qquad (1)$$

where $\vec{\alpha}$ and $\vec{\beta}$ are the control parameters; $\hat{D}(\vec{X})$ is the matrix of the diffusion coefficients; $\hat{A}(\vec{X})$ and $\hat{R}(\vec{X})$ are the stock and sinks functions.

In order to sustain permanent boundedness over the local accumulation of matter, the solution of eq.(1) must be cast in the form:

$$\vec{X}(\vec{r},t) = \int\int (X_{\omega,k}^{Re} + iX_{\omega,k}^{Im}) \exp(i\omega t + i\vec{k} \bullet \vec{r}) \qquad (2)$$

where the imaginary part $X_{\omega,k}^{Im}$ is the part of the solution $\vec{X}(\vec{r},t)$ supposed to provide emitting of the matter away from the given locality. The comparison between the real and imaginary parts of the l.h.s. and r.h.s. of eq.(1) sets that the conditions for $X_{\omega,k}^{Im}$ to be always non-zero are:

$$\omega + \vec{k} \bullet \nabla \hat{D}(\vec{X}) = 0$$
$$i\vec{\alpha} + i\vec{k} \bullet \hat{D}(\vec{X}) \bullet \vec{k} = 0 \qquad (3)$$

In order to make the calculations explicit, eq.(3) is carried out for $\hat{A}(\vec{X}) = \hat{1}$ and zero linear part of $\hat{R}(\vec{X})$. It is obvious, that the eq.(3) select unique non-zero values for $\omega$ and $\vec{k}$ which defines a wave that spreads with finite velocity. It should be stressed that the necessary condition for setting a well defined finite velocity of the wave is the boundedness of the local gradients (expressed through the gradient of the diffusion coefficient matrix $\nabla \hat{D}$). Note, that the fine-tuning of the matter wave to the control parameters is implemented through the dependence of the wave-vector on the control parameters $\vec{\alpha}$.

Outlining, the spontaneous emitting of a specific matter wave appears as a generic property of the reaction-diffusion systems coupled to mass/heat reservoirs whose aim is the avoidance of arbitrary accumulation of matter and energy in every locality.

Summarizing, the steady functioning of a network is grounded on its organization in cycles whose "homeostasis" is sustained through non-local autocatalytic feedbacks. In turn, it makes possible to associate semantic unit with each cycle and the coherence among functioning of distant cycles provides hierarchical order between different semantic units.

Let us now focus our attention the following aspect of the hierarchical organization of the semantic structure: so far we demonstrate how a sequence of letter turns to word by means of its representation through a performance of the corresponding engine; but how a word turns to syntactic unit on the next hierarchical level?! In the next subsection we shall demonstrate that this issue is interrelated with the issue of non-algorithmic computing.

## 2.3. Non-Algorithmic Computing

The organization into cycles viewed as necessary condition for steady functioning of a network prompts suggestion about the uniformity of the hierarchical structure of semantics - we assert that at any level of semantic organization a semantic unit is constituted by a specific sequence of "letters" such that its semantic meaning is represented by the specific performance of the corresponding engine. On the next hierarchical level the semantic unit as whole, represents syntactic unit ("letter") which must have an autonomous description irreducible to the sequence of letters on the original hierarchical level and its semantic meaning is to be associated with the performance of another engine. Next we shall consider how this goal is achieved for reaction-diffusion systems.

As already considered in sec. 1.3, a "letter" can be represented through any specific for the corresponding basin property. Recently we have proposed [13] that for the reaction-diffusion systems such property is the discrete band of the power spectrum of the solution of the corresponding system of differential equations (eq.(1)). Thus, each "letter" is characterized by the discrete band in its power spectrum. Further, the discrete band of the power spectrum of any sequence of letters (words) is non-linear combination of the discrete bands of the "letters". Thus, in turn the new "letter" (syntactic unit) appears as irreducible to the original sequence of "letters". Therefore, indeed, it is possible to achieve many-levels of hierarchical organization by means of following the above considered pattern.

It is worth noting, that the boundeness sets certain very important properties of the power spectrum of the reaction-diffusion systems. Indeed, the constraint over the amplitude of the terms in BIS puts a ban over possible resonances and thus sets that the difference between discrete bands of different letters to be substantiated only through presence of lines whose ratio is an irrational number. Physically this constraint is met by means of "driving" carried out by the spontaneously emitted material waves. This, in turn, makes "distances" between successive letters not achievable in an algorithmic way, i.e. they are not reducible to a finite sequence of arithmetic operations!

Now we shall demonstrate that the boundedness provides also reproducibility of the semantics under small fluctuations. Indeed, as proven in [14] the constraint of boundedness, viewed as avoiding resonances, sets additivity between the discrete band and the noise (continuous) band in the power spectrum of BIS. Note, that we consider BIS that are constituted under non-linear and non-uniform interplay between signal and noise. It is well known that in the general case, it is impossible to achieve a separation of the signal from the noise in the time series itself. Yet, it turns out, that on the contrary to the time series itself, it is possible to achieve such separation for the power spectrum of BIS - its power spectrum is additively decomposed to a specific discrete band and a continuous (noise) band whose shape is $1/f^{\alpha(f)}$. The invariance of this shape to the samples that represent the noise along with the additivity of the discrete and the continuous bands provides the reproducibility of the corresponding discrete band (letter).

Summarizing, the stabilizing role of the autocatalytic-type connections among nodes along with the reproducibility of the semantic units (discrete bands in the corresponding power spectra) makes semantic meaning to appear as a result of specific causal-like relations describing system behavior regardless to whether the system is a deterministic or a stochastic one.

**Conclusions**

It has been demonstrate that under the hypothesis of boundedness the semantics appears as a property of spontaneous physical processes. The boundedness sets an exclusive two-fold representation of a semantic unit: as a specific sequence of letters and as a performance of a specific engine so that their interplay serves as grounds for building a multi-layer hierarchy of semantic structures. A radically new property of this organization is that the semantics is associated with the steady processes whose behavior is described by causal-like relations, though the processes themselves might be stochastic. This result determines unambiguous relation between causality and semantics.